\documentstyle[prl,aps,epsfig,twocolumn,amssymb]{revtex}

\begin{document}
\newcommand{\beq}{\begin{eqnarray}}
\newcommand{\eeq}{\end{eqnarray}}
\newcommand{\non}{\nonumber\\}
\newcommand{\y}{{\bf Y}}
\newcommand{\A}{{\bf A}}
\newcommand{\M}{{\bf M}}
\newcommand{\m}{{\bf m}}
\newcommand{\U}{{\bf U}}
\newcommand{\V}{{\bf V}}
\newcommand{\N}{{\bf N}}
\newcommand{\tl}{{\widetilde{L}}}
\newcommand{\tr}{{\widetilde{R}}}
\newcommand{\tn}{{\tilde{\nu}}}
\newcommand{\tm}{{\widetilde{m}}}
\newcommand{\tll}{{\tilde{\ell}}}
\newcommand{\yudyu}{{\bf Y}_U^\dag{\bf Y}_U}
\newcommand{\yddyd}{{\bf Y}_D^\dag{\bf Y}_D}
\newcommand{\yldyl}{{\bf Y}_\ell^\dag{\bf Y}_\ell}
\newcommand{\yndyn}{{\bf Y}_\nu^\dag{\bf Y}_\nu}
\newcommand{\aldal}{{\bf A}_\ell^\dag{\bf A}_\ell}
\newcommand{\andan}{{\bf A}_\nu^\dag{\bf A}_\nu}
\newcommand{\tchi}{\tilde{\chi}}
\newcommand{\gab}{g_{\alpha\beta}}
\draft
\title{Lepton flavor violating $Z$-decays in supersymmetric seesaw model}
\author{Junjie Cao $^{1,2}$, Zhaohua Xiong $^{3}$, Jin Min Yang $^2$}

\address{$^1$ CCAST (World  Laboratory), P.O.Box 8730, Beijing 100080, China}
\address{$^2$ Institute of Theoretical Physics, Academia Sinica, Beijing
              100080, China}
\address{$^3$ Graduate School of Science, Hiroshima University,
              Hiroshima 937-6256, Japan}

\date{\today}

\maketitle
\begin{abstract}

In supersymmetric seesaw model, the large flavor mixings of sleptons 
induce the lepton flavor violating (LFV) interactions
$\ell_I \bar\ell_J V$ ($V=\gamma, Z$), which give rise to various
LFV processes. In this work we examine the LFV decays
$Z\to\ell_I \bar\ell_J$. Subject to the constraints from the
existing neutrino oscillation data and the experimental bounds on
the decays $\ell_J\to\ell_I\gamma$, these LFV $Z$-decays are found
to be sizable, among which the largest-rate channel $Z\to
\tau \bar{\mu}$ can occur with a branching ratio of $10^{-8}$ 
and may be accessible at the LHC or GigaZ experiment.
\end{abstract}


\section{Introduction}
\label{Sec:intro}

It is well known that the Standard Model (SM) predicts an
unobservably small branching ratio for any lepton flavor violating
(LFV) process, such as  $\ell_J\to\ell_I\gamma$ or
$Z\to\ell_I\bar\ell_J$. In some extensions of the SM the LFV
processes may be significantly
enhanced~\cite{Casas01,Hisano02,Ellis00}. One example of these
extensions is the popular weak-scale supersymmetry (SUSY). In SUSY
models the LFV interactions  $\ell_I \bar\ell_J V$ ($V=\gamma, Z$)
\cite{Frank96,Carvalho02,Atwood02,Illana02} receive two kinds of
additional loop contributions: One is from the charged-current
lepton-sneutrino-chargino couplings; the other is from the flavor
mixings of charged sleptons.  While the former is a common feature
of all SUSY models accommodating right-handed neutrinos, the
latter is sizable only in some specific realizations of SUSY,
such as the minimal supergravity model (mSUGRA) \cite{mSUGRA} 
with seesaw mechanism to generate the tiny masses for light
neutrinos.  The mechanism is realized by introducing  
right-handed neutrino superfields~\cite{Casas01,Hisano02} with
very heavy Majorana masses. 
In such a framework the flavor diagonality of charged sleptons is 
usually assumed at the Planck
scale, but the flavor mixings at weak scale are inevitably
generated through renormalization equations since there is no
symmetry to protect the flavor diagonality. Such 
flavor mixings of charged sleptons generated at weak scale are 
proportional to neutrino Yukawa coupling, which may be as large as 
top quark Yukawa coupling due to seesaw mechanism, and are enhanced 
by a large factor $\log(M_P^2/{\cal M}^2)$ ($M_P$ is Planck scale and 
${\cal M}$ 
is the neutrino Majorana mass). Therefore, the popular mSUGRA with seesaw
mechanism predicts large flavor mixings of sleptons at weak scale,
which will reveal their effects through some LFV processes in
collider experiments.

The aim of this article is to examine the LFV $Z$-decays 
$Z\to\ell_I \bar\ell_J$
induced by slepton flavor mixings in the mSUGRA seesaw model.
Given the possibility of the extremely accurate measurement of $Z$-decays
in future experiments, the decays $Z\to\ell_I \bar\ell_J$ may serve
as a sensitive probe for such a new physics model.

We will use the existing neutrino oscillation data
and the experimental bounds on the decay $\ell_J\to\ell_I\gamma$
to constrain the model parameters, and then evaluate the
branching ratios of $Z\to\ell_I \bar\ell_J$.  We find that, subject to
the current constraints, the channel $Z\to \tau \bar{\mu}$ can
occur with a branching ratio as large as $10^{-8}$ and
thus may be accessible at LHC  \cite{Martin00}
or the GigaZ option of TESLA at DESY \cite{AS01}.

This article is organized as follows. In section II, we briefly
describe the SUSY seesaw model with {\em minimal} CP-violation in
the right-hand neutrino sector and discuss the induced flavor
mixings between sleptons. In section III, we present the analytic
results for the SUSY contributions to the branching ratio of
$Z\to\ell_I\bar{\ell}_J$. In section IV, we present the
correlation between the process $Z\to \ell_I\bar{\ell_J}$
 and $\ell_J\to \ell_I\gamma$.
 In section V, we evaluate the numerical size of the branching ratio of
$Z\to\ell_I\bar{\ell}_J$. Finally in section VI, we
give our conclusion.

\section{Supersymmetric seesaw model and charged slepton mixings }
\subsection{Supersymmetric seesaw model}

The seesaw mechanism~\cite{Yanagida80} provides an elegant
explanation for the tiny masses of light neutrinos, which
implies that new physics scale is about $10^{14}$ GeV.
However, a non-symmetric seesaw model suffers from a serious 
hierarchy problem \cite{Casas01}, which can be  automatically solved
in the SUSY framework.

In supersymmetric seesaw model with $N$ right-handed neutrino
singlet fields $\nu_R$, additional terms in the superpotential
arise~\cite{Casas01}: 
\beq 
W_\nu = -\frac{1}{2}\nu_R^{cT}\M
\nu_R^c + \nu_R^{cT} \y_\nu L \cdot H_2 \ , \label{eq1} 
\eeq 
where
$\M$ is $N\times N$  mass matrix for the right-handed neutrino,
and $L$ and $H_2$ denote the left-handed lepton and the Higgs
doublet with hypercharge $-1$ and $+1$, respectively. At energies
much below the mass scale of the right-handed neutrinos, the
superpotential leads to the following mass matrix for the
left-handed neutrinos: 
\beq 
\M_\nu = \m_D^T {\M}^{-1}\m_D =
\y_\nu^T {\M}^{-1}\y_\nu (v\sin\beta)^2 \ . \label{nrmass} 
\eeq
Obviously, the neutrino masses tend to be light if the mass scale
${\cal M}$ of the matrix $\M$ is much larger than the scale of the
Dirac mass matrix $\m_D=\y_\nu \langle H_2^0\rangle=\y_\nu
v\sin\beta$ with $v=174$ GeV and $\tan\beta =\langle
H_2^0\rangle/\langle H_1^0\rangle$. The matrix $\M_{\nu}$ can be
diagonalized by the MNS  matrix $\U_\nu$: 
\beq
\U_\nu^\dagger\M_\nu\U^*_\nu=diag(m_{\nu 1}, m_{\nu 2}, m_{\nu 3})\ ,
\label{NeutrinoDiag} 
\eeq 
where $m_{\nu i}$ are the light neutrino masses.
\subsection{Slepton flavor mixings}
The mass matrix of the charged sleptons is given by
\beq
\m_{\tilde \ell}^2=\left(
    \begin{array}{cc}
        \m_{\tll LL}^2    &\m_{\tll LR}^{2\dag} \\
        \m_{\tll LR}^2   &\m_{\tll RR}^2
    \end{array}
      \right)
\label{slepmass}
\eeq
with
\beq
\m^2_{\tll LL}&=&\m_\tl^2+\left[m_{\ell}^2+m_Z^2
\left(-\frac{1}{2}+s^2_W\right)\cos 2\beta \right]{\Large\bf 1} ,
\label{mlcharged} \\
\m^2_{\tll RR}&=&\m_\tr^2+\left(m_{\ell}^2
- m_Z^2 s^2_W\cos 2\beta \right){\Large\bf 1} ,\\
\m^{2}_{\tll LR} &=&\A_\ell v\cos\beta-m_\ell\mu\tan\beta ~{\Large\bf 1} ,
\eeq
where $s_W=\sin\theta_W$, $c_W=\cos\theta_W$, 
$\theta_W$ is the Weinberg angle and ${\Large\bf 1}$ is unit
$3\times 3$ matrix in generation space.
In mSUGRA model it is assumed that at the Planck scale
the soft-breaking parameters satisfy
\beq
\m_\tl&=&\m_\tr=\m_\tn=m_0{\bf\Large 1},\ \  m_{H_1}=m_{H_2}=m_0,\non 
 \A_\ell&=&A_0\y_\ell, \ \ \A_\nu=A_0\y_\nu.
\eeq 
In general, the lepton Yukawa couplings $\y_\ell$ and
$\y_\nu$ cannot be diagonalized simultaneously. It is usually
assumed that $\y_\ell$ is flavor diagonal but  $\y_\nu$ is not. In
this basis the mass matrix of the charged sleptons is flavor
diagonal at Planck scale. However, when evolving down through
renormalization group (RG) equations (see Appendix A) to weak scale, such
flavor diagonality is broken.  In the leading-log approximation,
we have~\cite{Hisano02} 
\beq \label{eq:rnrges} \label{dmLij}
\delta (\m_{\tilde L}^2)_{IJ} &\simeq&
        -\frac{1}{8\pi^2}(3m_0^2+A_0^2)(\y_\nu^{0\dag}\y_\nu^0)_{IJ}
        \ln\left(\frac{M_P}{\cal M}\right) \ , \\
\delta(\m_{\tilde R}^2)_{IJ} &\simeq & 0 \ ,\\
  \label{smlreq}
\delta (\A_\ell)_{IJ} &\simeq& -\frac{3}{16\pi^2}A_0(\y^0_\ell)_{II}
      (\y_\nu^{0\dag}\y_\nu^0)_{IJ} \ln\left(\frac{M_P}{\cal M}\right)\ ,
\eeq
where $\y^0\equiv \y(M_P)$.

The flavor non-diagonal mass matrix $\m^2_\tll$ in
Eq.(\ref{slepmass}) at weak scale can be diagonalized by a unitary
matrix ${\bf S}_\tll$ 
\beq 
{\bf S}_\tll\m^2_\tll{\bf S}_\tll^\dag =diag(m^2_{\tll_X}). 
\eeq 
Such a unitary rotation of slepton
fields is to induce the flavor-changing  neutral-current vertices:
$\tilde \chi^0_\alpha \ell_I \tilde\ell_X$ and $Z \tilde\ell_X
\tilde\ell_Y$.

In supersymmetric seesaw model, there exist right-handed
sneutrinos with the same order masses as  the heavy Majorana
neutrinos. However, due to their large masses, they do not give
significant contributions to the considered LFV processes.
Therefore, only the left-handed sneutrinos need to be take into
account, whose mass matrix is given by 
\beq
\m^2_{\tilde{\nu}}=\m^2_\tl+\frac{1}{2}m_Z^2
       \cos 2\beta\ {\bf\Large 1} .
\eeq 
Due to the non-diagonal contribution $\delta (\m_{\tilde
L}^2)_{IJ}$ in Eq.(\ref {dmLij}),  $\m^2_{\tilde{\nu}}$ is flavor
non-diagonal at weak scale and needs to be diagonalized by a
unitary matrix ${\bf S}_{\tilde{\nu}}$ 
\beq 
{\bf S}_{\tilde{\nu}}\m^2_{\tilde{\nu}}{\bf S}_{\tilde{\nu}}^\dag
     =diag(m^2_{\tilde{\nu}_X}). 
\eeq 
Such a unitary rotation of
sneutrino fields results in the charged-current flavor-changing
vertex: $\tilde \chi^+_\alpha \ell_I \tilde\nu_X$.

\subsection{The form of neutrino Yukawa coupling}
As shown in Eqs.(\ref{eq:rnrges}) and (\ref{smlreq}), the flavor
mixings of charged sleptons are proportional to neutrino Yukawa
couplings. Lack of knowledge of the neutrino Yukawa couplings
results in numerous speculations on their  possible forms. 
Different forms may lead to different  flavor
mixings.
In this work we consider a scenario called as {\em minimal} CP violating
seesaw model which has two heavy Majorana neutrinos with the
Dirac mass matrix $\m_D$ parameterized as \cite{Endoh02} 
\beq
 \m_D^T \equiv \y_\nu^T \langle H_2^0\rangle =\U_L{\bf m}{\bf V}_R,\ \
{\bf m}=\left(\begin{array}{cc}
     0 & 0\\
     m_2 & 0\\
     0 & m_3
     \end{array}\right),
\label{lnumass} \eeq where  \beq {\bf V}_R=\left(\begin{array}{cc}
      \cos\theta_R& \sin\theta_R\\
     -\sin\theta_R&\cos\theta_R
     \end{array}\right)
     \left(\begin{array}{cc}
      e^{-i\gamma_R/2}& 0\\
     0&e^{i\gamma_R/2}
     \end{array}\right), \label{vr}
\eeq 
with mixing angle $\theta_R$ and CP violating phase
$\gamma_R$ for the heavy Majorana neutrinos concerning directly
with leptogenesis \cite{Endoh02}. The matrix $\U_L$ appearing in
Eq.~(\ref{lnumass}) reads 
\small
\beq 
\U_L={\bf
O}_{23}(\theta_{L23})\U_{12}(\theta_{L13}, \delta_L) {\bf
O}_{12}(\theta_{L12}) {\bf P}(-\gamma_L/2) \ , \label{ul}
\eeq 
\normalsize
where ${\bf P}(-\gamma_L/2)={\rm diag}[1, {\rm exp}(-i\gamma_L/2),
{\rm exp}(i\gamma_L/2)]$, and ${\bf O}_{ij}$ and $\U_{ij}$ denote the
rotations in $(i,j)$ plane. Without loss of generality, $m_{2,3}$
in Eq.(\ref{lnumass}) are chosen to be real, positive and
$m_2<m_3$.

As Eqs.\ (\ref{nrmass}) and (\ref{lnumass}) are used, 
the mass matrix for the
light neutrinos in this model can be further expressed as 
\beq
\M_{\nu} = \U_L \m \V_R \M^{-1} \V_R^T \m^T \U_L^T. 
\eeq 
And the MNS matrix in (\ref{NeutrinoDiag}) is found to be 
a product of matrices,
\beq
\U_{\nu}=\U_L {\bf K}_R, 
\eeq   
where ${\bf K}_R ={\bf K}_R (\theta, \phi, \alpha) $ is a unitary
matrix. Therefore, now Eq.\ (\ref{NeutrinoDiag}) can be rewritten as 
\beq 
{\bf K}_R^{\dag} \m \V_R \M^{-1} \V_R^T \m^T {\bf K}_R^{\ast} ={\rm diag}[m_{\nu_1},m_{\nu_2},m_{\nu_3}].
\eeq 
From this equation, one can learn both ${\bf K}_R $ and $m_{\nu_i}$ 
are independent of the choice of $\U_L$.

It is noticeable that the special form (\ref{lnumass}) for
the neutrino Yukawa couplings matrix $\y_\nu$ implies\cite{Endoh02}:\\
(1) One of the neutrinos is massless, i.e., $m_{\nu 1}=0$.\\
(2) The quantity $\yndyn$ is only dependent on 3 mixing angles
$\theta_{L12, L13, L23}$ and a CP violating 
phase $\delta_L$ in ${\bf U}_L$, 
\small 
\beq
 (\yndyn)_{IJ}=\frac{m_2^2(\U_L)_{I2}(\U_L^\dag)_{2J}+
m_3^2(\U_L)_{I3}(\U_L^\dag)_{3J}}{(v\sin\beta)^2}.
\label{yy}
\eeq
\normalsize
(3) For small mixing angles $\theta_{L13}$ and $\theta$,
               the light neutrinos mixing matrix $\U_\nu$ takes
               a simplified form similar to the mixing matrix
               introduced in \cite{Maki62}:
\small
\beq
\U_\nu\simeq \left(\begin{array}{ccc}
         c_{L12}&s_{L12}& \begin{array}{l}s_{L13}e^{-i\delta_L}\\
                          +s_{L12}s_\theta e^{-i\phi'} \end{array}\\
        -s_{L12}c_{L23}&c_{L12}c_{L23}&s_{L23}\\
        s_{L12}s_{L23}&-c_{L12}s_{L23}&c_{L23}
         \end{array}\right){\bf P}(\alpha')
\label{UMNS}
\eeq
\normalsize
where $\phi'=\phi+\gamma_L$, $\alpha'=\alpha-\gamma_L/2$ and
$s_x\equiv\sin x$, $c_x\equiv\cos x$.
In this case, the angles in $\yndyn$ can be related directly to
the corresponding neutrino mixing angles and determined by
neutrino experiments.
\section{The LFV decays $Z\to\ell_I\bar{\ell}_J$}
\label{Sec:zdecay}

\begin{figure}[hbt]
\epsfig{file=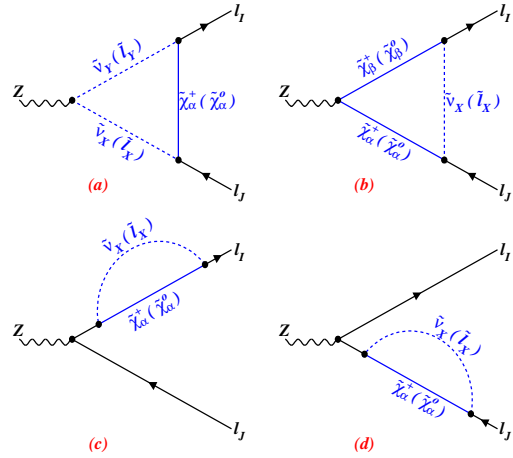,width=7cm} \vspace*{0.3cm} \caption{Feynman
diagrams of SUSY contributions to the LFV processes
$Z\to\ell_I\bar{\ell}_J$. \label{Fig1}}
\end{figure}

\begin{figure}[htb]
\epsfig{file=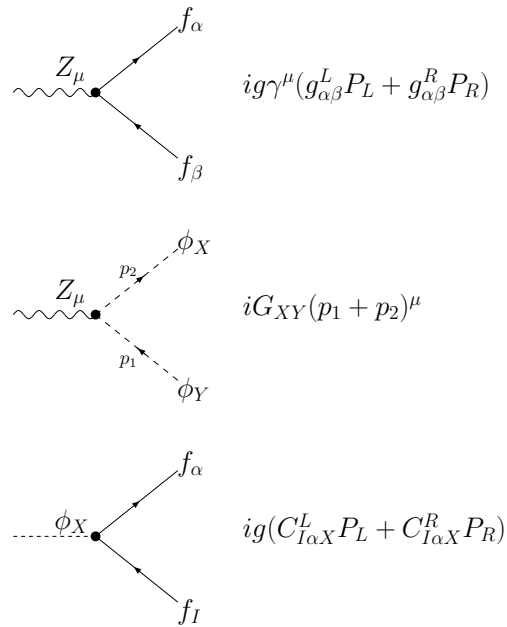,width=7cm} \caption{Some interaction vertices
needed to  calculate the branching ratio of
$Z\to\ell_I\bar{\ell}_J$ in SUSY. $\alpha$ and $\beta$ are indices
of charginos (neutralinos), while $X$ and $Y$ are those for
sleptons. \label{Fig2}}
\end{figure}

The flavor changing interactions in slepton sector discussed in
the preceding section, namely the couplings $\tilde \chi^0_\alpha
\ell_I \tilde\ell_J$ and $Z \tilde\ell_I \tilde\ell_J$ from
charged slepton mixings as well as $\tilde \chi^+_\alpha \ell_I
\tilde\nu_J$ from sneutrino mixings, can induce the LFV processes
$Z\to\ell_I\bar{\ell}_J$, as shown in Fig.~\ref{Fig1}. The
relevant Feynman rules can be derived straightforwardly from the
analysis in the preceding section. Our analytic results will be
expressed in terms of the constants $\gab^{L,R},\ G_{XY}$ and
$C_{I\alpha X}^{L,R}$ defined in Fig.~\ref{Fig2}, whose explicit
expressions can be found in \cite{Illana02,Haber85}.
The calculation of the diagrams in Fig.~\ref{Fig1} results in an
effective $Z\bar{\ell}_I \ell_J$ vertex: 
\beq {\cal M}=ig\epsilon_\mu\bar{u}_{\ell_I}(p_1)\Gamma^\mu u_{\ell_J}(p_2)
\label{ampZ} 
\eeq 
with $\varepsilon_\mu$ being the polarization
vector of $Z$-boson, $p_1$($p_2$) the momentum of
$\ell_I$($\ell_J$), and $\Gamma^\mu$ given by 
\beq
\Gamma^\mu&=&\frac{\alpha_{em}}{\sin^2\theta_W} \left[
\gamma^\mu\left(f_{1L}P_L+f_{1R}P_R\right)\right.\non
&&\left.+i\sigma^{\mu\nu}k_\nu
 \left(f_{2L}P_L+f_{2R}P_R\right)\right] \ ,
\label{factors} 
\eeq 
where $P_{R,L}=\frac{1}{2}(1\pm \gamma_5)$,
$g=e/\sin\theta_W$ and $k=p_1-p_2$ is the momentum transfer. The
form factors $f_{1L}$, $f_{1R}$, $f_{2L}$ and $f_{2R}$ arising
from the calculation of the loop diagrams in Fig.~\ref{Fig1} are
listed as
follows.\\
\underline{Contribution of Fig.~\ref{Fig1}(a)}:
\beq
f_{1L}^a&=&G_{XY}C_{I\alpha X}^{*L}\left[-2C^a_{24}C_{J\alpha Y}^{L} \right.\non
 & & \left.+m_{\ell_J} m_\alpha\left(C_0^a+C_{11}^a+C_{12}^a\right) C_{J\alpha Y}^{L}\right], \label{f1la} \\
f_{2L}^a &=& G_{XY}C_{I\alpha X}^{*R}\left[
  m_\alpha \left(C_0^a+C_{11}^a+C_{12}^a\right)C_{J\alpha Y}^{L}\right.\non
& & \left.-m_{\ell_J}\left(C_{12}^a +C_{22}^a+C_{23}^a\right)
C_{J\alpha Y}^{R}\right] \ .
\eeq 
\underline{Contribution of Fig.~\ref{Fig1}(b)}: 
\beq 
f_{1L}^b&=&C_{I\alpha X}^{*L}C_{J\beta
X}^L\left[ \gab^L m_\alpha m_\beta C_0^b\right.\non
&&\left.+\gab^R\left(m_Z^2C_{23}^b-2C_{24}^b+\frac{1}{2}\right)\right]\non
&&+C_{I\alpha X}^{*L}C_{J\beta X}^R
\gab^L m_\alpha m_{\ell_J} \left(C_0^b+C_{11}^b+C_{12}^b\right), \\
f_{2L}^b&=&C_{I\alpha X}^{*R}C_{J\beta X}^{L}
\left(\gab^R m_\alpha C_{11}^b+\gab^Lm_\beta C_{12}^b\right)\non
&&+C_{I\alpha X}^{*R}C_{J\beta X}^R
\gab^L m_{\ell_J}\left(C_{12}^b+C_{22}^b+C_{23}^b\right).
\label{f1lb}
\eeq
\underline{Contribution of Fig.~\ref{Fig1}(c) plus Fig.~\ref{Fig1}(d)}:
\beq
f_{1L}^{c}&=&C_{I\alpha X}^{*L}\left[\frac{m_\alpha}{m_J}(B_0^1-B_0^2)
C_{J\alpha X}^{*R}-B_1^1C_{J\alpha X}^{L}\right]g_L,\\
f_{1R}^{c}&=&0 \ . \label{f1lc} 
\eeq 
In the above, $g_L=(1-2\sin^2\theta_W)/(2 \cos\theta_W)$,  and
$B_{0,1}^i=B(-p_i; m_\alpha^2,m_X)$,
$C_{0,ij}^a=C_{0,ij}\left(-p_1,-p_2;
m_\alpha^2,m_Y^2,m_X^2\right)$ and
$C_{0,ij}^b=C_{0,ij}\left(-p_1,-p_2;
m_X^2,m_\beta^2,m_\alpha^2\right)$ are the Feynman loop integral
functions~\cite{Hooft79}. Terms proportional to the lepton masses
$m_{\ell_I}$ are neglected. The right-handed form factors from the
vertex loops are obtained from the corresponding left-handed ones
in (\ref{f1la})-(\ref{f1lb}) by the substitution $L\leftrightarrow
R$.

The branching ratio of $Z\to \ell_I\bar\ell_J$ (including its
charge-conjugate channel) is then given by \footnote{Our result is
in agreement with that given in \cite{Illana02} 
if $m_{\ell_J}$-dependence terms in $f_{1L,1R}$ are neglected.} 
\beq 
{\cal B}r(Z\to \ell_I \bar\ell_J)&=&
\frac{1}{48\pi^2}\left(\frac{\alpha_{em}}{\sin^2\theta_W}\right)^3
\frac{m_Z}{\Gamma_Z} \left[|f_{1L}|^2+|f_{1R}|^2\right.\non
&&\left.+\frac{m_Z^2}{2}\left( |f_{2L}|^2+|f_{2R}|^2\right)\right]
\ , \label{brzij0} 
\eeq 
where
$f_{iL,iR}=\sum\limits_{\alpha=a,b,c}f_{iL,iR}^\alpha$ and
$\Gamma_Z$ denotes the total decay width of Z boson.

Although the above results are sufficient to allow for numerical 
calculations, we would like to derive an analytical expression
for the branching ratio by  
considering the limit $m_S\gg m_Z$ where $m_S$ represents the mass of any
internal sparticle in the loops in Fig.\ref{Fig1}. 
In this case the loop functions can be much simplified and
we use the mass-insertion approximation in our derivation. 
In such a limit, the chargino mass matrix
\beq
     \M_{\tchi^\pm}=\left(\begin{array}{cc}
       M_2  & \sqrt{2}m_W\sin\beta\\
       \sqrt{2}m_W\cos\beta & \mu
       \end{array}\right)
\eeq
is nearly diagonal. Here  $\mu$ is the mass parameter appearing in
the term $\mu H_1 H_2$ in superpotential and $M_2$ is the $SU(2)$
gaugino mass parameter. The matrices ${\bf U}$ and ${\bf V}$ which diagonalize
$\M_{\tchi^\pm}$ will be unit ones for $\mu>0$, and the chargino
masses are given by 
\beq
 m_{\tchi_1^\pm}=M_2, \ \ m_{\tchi_2^\pm}=|\mu|.
\eeq 
The symmetric neutralino mass matrix
   \beq
     \M_{\tchi^0}=\left(\begin{array}{cccc}
       M_1  &    & &  \\
       0    & M_2 & &  \\
       -m_Zs_Wc_\beta & m_Zc_Wc_\beta &0 & \\
       m_Zs_Ws_\beta & -m_Zc_Ws_\beta &-\mu &0
       \end{array}\right)
       \eeq
can be diagonalized by a unitary matrix ${\bf N}$
   \beq
     {\bf N}=\left(\begin{array}{cccc}
       1  &  & & \\
        &1   & &  \\
        & &\sqrt{2}e^{i\frac{\pi}{4}}&-\sqrt{2}e^{-i\frac{\pi}{4}}\\
        & &-\sqrt{2}e^{-i\frac{\pi}{4}}&\sqrt{2}e^{i\frac{\pi}{4}}
       \end{array}\right).
       \eeq
The corresponding  mass eigenvalues are given by 
\beq
 m_{\tchi_{1,2}^0}=M_{1,2},\ \ \
 m_{\tchi_3^0}=m_{\tchi_4^0}=|\mu|.
\eeq 
When using mass-insertion method, one should note the fact
that, for any matrix $\M=\M^0+\M^1$, where
$\M^0={\rm diag}(m_1^0,\cdots, m_n^0)$ and $\M^1$ has no diagonal
elements, if matrix ${\bf T}$ can diagonalize the matrix $\M$,
${\bf T M T}^\dag={\rm diag}(m_1,m_2,\cdots, m_n)$, then at leading
order for an arbitrary function $f$ 
\beq 
{\bf T}_{ik}^\dag
f(m_k){\bf T}_{kj}= \delta_{ij}f(m_i^0)+\M^1_{ij}f(m_i^0, m_j^0)
\eeq 
with 
\beq 
f(x,y,z_1\cdots z_n)= \frac{f(x,z_1 \cdots
     z_n)-f(y,z_1\cdots z_n)}{x-y}. \label{func} 
\eeq
After a straightforward calculation we obtain an 
analytical expression for the branching ratio 
\small 
\beq {\cal B}r(Z\to \ell_I\bar\ell_J)
&=&\frac{\alpha^3_{em}}{48\pi^2}\frac{c^2_W}{s^6_W}
\frac{m_Z}{\Gamma_Z}\frac{|\delta
(m^2_\tl)_{IJ}|^2}{M_2^4}\left|f_1(x_I,x_J) \right.\non
&&\left.-2f_2(x_I,x_J)-\frac{\frac{1}{2}+s^2_W}{c^2_W}
\frac{f_2(\frac{1}{x_I},\frac{1}{x_J})}{x_Ix_J}\right.\non
&&\left.+\frac{\frac{1}{2}s^2_W-s^4_W}{c^4_W}
\left(\frac{M_2}{M_1}\right)^2
\left(\frac{f_2(\frac{1}{x^\prime_I},\frac{1}{x^\prime_J})}
{x^\prime_Ix^\prime_J}\right.\right.\non
&&\left.\left.-\frac{1}{2}f_3(x^\prime_I,x^\prime_J)\right)
-\frac{3}{2}\frac{\frac{1}{2}-s^2_W}{c^2_W} f_3(x_I,x_J)\right|^2
. \label{brzij} 
\eeq 
\normalsize 
Here $s_W=\sin\theta_W$,
$c_W=\cos\theta_W$, $x_I=(m^2_\tl)_{II}/M_2^2$,
$x^\prime_I=(m^2_\tl)_{II}/M_1^2$, and 
\beq
f_1(x)&=&\frac{1}{x-1}\left(1-\frac{x}{x-1}\ln x\right),\non
f_2(x)&=&\frac{1}{4(x-1)}\left(1-\frac{x^2}{x-1}\ln x\right),\non
f_3(x)&=&\frac{1}{(x-1)}\left(1+\frac{x^2-2x}{x-1}\ln x\right),
\eeq 
and $f_i(x,y)$ can be obtained through Eq.~(\ref{func}).

\section{Comparison of LFV $Z$-decays with lepton decays}

Now we compare the LFV $Z$-decays with lepton decays. Using a
similar procedure in the preceding section, we can  easily
calculate the decay width for $\ell_J\to\ell_I\gamma$ by setting
$g=e$, $g_{\alpha\beta}^{L,R}=1$ with $\alpha=\beta$, $G_{XY}=1$
with $X=Y$ in Fig.\ \ref{Fig2}, and $f_{1L,1R}=0$
in (\ref{factors}). Meanwhile one should also note the fact that
sneutrinos in Fig.\ \ref{Fig1}(a) and neutralinos in \ref{Fig1}(b)
do not couple to photon and that the self-energy diagrams do not
contribute to dipole operators. The branching ratios of
$\ell_J\to\ell_I\gamma$ are obtained as 
\beq 
\frac{{\cal B}r(\ell_J\to\ell_I\gamma)} {{\cal
B}r(\ell_J\to\ell_I\nu_J\bar{\nu}_I)}=
\frac{6\alpha_{em}}{\pi}\frac{m_W^4}{m^2_{\ell_J}}
\left(|f^\gamma_{2L}|^2+|f^\gamma_{2R}|^2\right) . \label{dwij0}
\eeq Here the form factors are given by \cite{Illana02} \small
\beq f^\gamma_{2L}&=&\sum\limits_{k=a,b}\frac{1}{m^2_\alpha}
C_{I\alpha X}^{*R(k)} \left[m_\alpha C_{J\alpha X}^{L(k)}F_1^k
+m_{\ell_J}C_{J\alpha X}^{R(k)}F_2^k\right] ,\\ \label{factorg}
f^\gamma_{2R}&=&f^\gamma_{2L}\left \vert_{L\leftrightarrow R} \right. ,
\eeq
\normalsize
where
\small
\beq
F_1^a(x_a)&=&m^2_{{\tilde\chi}_\alpha^0}(C_0^a+C_{11}^a+C_{12}^a)\non
     &=&\frac{1}{(x_a-1)^2}\left(-\frac{x_a+1}{2}
    +\frac{x_a}{x_a-1}\ln x_a\right), \\
F_2^a(x_a)&=&-m^2_{{\tilde\chi}_\alpha^0}(C_{12}^a+C_{22}^a+C_{23}^a)\non
     &=&\frac{1}{2(x_a-1)^3}\left(\frac{-x_a^2+5x_a+2}{6}-\frac{x_a}{x_a-1}\ln
     x_a\right), \\
F_1^b(x_b)&=&m^2_{{\tilde\chi}_\alpha^-}(C_{11}^b+C_{12}^b)\non
     &=&\frac{1}{(x_b-1)^2}
\left(\frac{-3x_b+1}{2}+\frac{x_b^2}{x_b-1}\ln x_b\right), \\
F_2^b(x_b)&=&m^2_{{\tilde\chi}_\alpha^-}
(C_{12}^b+C_{22}^b+C_{23}^b)
=\frac{1}{x_b}F_2^a(\frac{1}{x_b})
\eeq
\normalsize
with $x_a=m^2_{{\tilde\ell}_X}/m^2_{{\tilde\chi}_\alpha^0}$ and
$x_b=m^2_{{\tilde\nu}_X}/m^2_{{\tilde\chi}_\alpha^-}$.

Next we derive the analytical expression for the branching ratios
in the limit of $m_S>>m_Z$. Unlike the form factors for the
$Z$-decays which contain terms not proportional to the small
lepton mass [see Eqs.\ (\ref{f1la}) and (\ref{f1lb})], the form
factors for $\ell_J\to\ell_I\gamma$ are always proportional to the
small lepton mass $m_{\ell_J}$. In this case, the off-diagonal
elements in the mass matrices of chargino and neutralino are no
longer negligible, especially when $\tan\beta$ is large. In fact,
the terms $m_\alpha C_{I\alpha X}^{*L(b)}C_{J\alpha X}^{R(b)}$ in
$f_{2R}^\gamma$ receive the contribution from the wino-Higgsino
mixing, which can be enhanced by $\tan\beta$. So for a large
$\tan\beta$, the contribution of $f_{2R}^\gamma$ is dominant and
the branching ratios are given by \cite{Hisano02} 
\beq {\cal B}r(\ell_J\to\ell_I\gamma)&\simeq& {\cal B}r(\ell_J\to\ell_I\nu_J\bar{\nu}_I)
\frac{6\alpha_{em}}{\pi}\frac{m_W^4}{m^2_{\ell_J}}
|f^\gamma_{2R}|^2\non &=&{\cal
B}r(\ell_J\to\ell_I\nu_J\bar{\nu}_I)
\frac{6\alpha_{em}}{\pi}\frac{m^4_W}{M_2^4}\left(\frac{\mu}{M_2}\right)^2\non
&\times&\left|\frac{1}{2}F^a_1(x_I,x_J)-F^b_1(x_I,x_J)\right.\non
&-&\left.\left(\frac{M_2}{\mu}\right)^4
\left(\frac{1}{2}F^a_1(\bar{x}_I,\bar{x}_J)
-F^b_1(\bar{x}_I,\bar{x}_J)\right)\right|^2\non
&\times&\frac{|\delta(\m^2_{\tilde L})_{IJ}|^2}{M_2^4}
\frac{\tan^2\beta}{(1-\frac{\mu^2}{M^2_2})^2} , \label{dwij} 
\eeq
where $\bar{x}_I=(\m^2_{\tilde L})_{II}/\mu^2$.

Comparing ${\cal B}r(\ell_J\to \ell_I\gamma)$ with ${\cal B}r(Z\to \ell_I\bar \ell_J)$, we find:

 (1) The dipole transitions in (\ref{factors}), the only operators
     contributing to $\ell_J\to\ell_I\gamma$, do not give dominate
     contributions to decays $Z\to \ell_I\bar{\ell}_J$ due to heavy sparticle
      mass suppression;

  (2) ${\cal B}r(Z\to \ell_I\bar \ell_J)$ is not sensitive to
      $\tan\beta$, whereas  ${\cal B}r(\ell_J\to \ell_I\gamma)$ can be
      enhanced by large $\tan\beta$;

   (3)  The ratio ${\cal B}r(Z\to \ell_I\bar{\ell}_J)/{\cal B}r(\ell_J\to \ell_I\gamma)$ 
        is independent of the heavy Majorana
       sector introduced by seesaw mechanism.

\section{Numerical results}

In our numerical calculation we consider the constraints
from current neutrino oscillation  experiments and the
experimental bounds on LFV lepton decays.

{\em (1) Neutrino oscillation  experiments:~}\\
The SK Collaboration~\cite{SK} showed that the $\nu_\mu$ created
in the atmosphere oscillates into $\nu_\tau$ with
almost maximal mixing, $\sin(2\theta_{atm})\sim 1$
and the neutrino mass-square difference is
$\Delta m^2_{atm}\sim (2-4) \times 10^{-3}$ eV$^2$.
The second  mass-square difference and mixing angle
are found to be $\Delta m^2_{sol}=(3-15)\times 10^{-5}$ eV$^2$,
 $\sin(2\theta_{sol})=0.7\sim 0.9$ from solar neutrino experiments\cite{SNO,KM}.
For the third mixing angle, only the upper bound is obtained from
the reactor neutrino experiments \cite{CHOOZ,PaloVarde}:
$\sin^22\theta_{rea} < 0.1$ for $\Delta m^2_{atm} \simeq 3\times 10^{-3}$ eV$^2$.

Although there exists a possibility that neutrino
masses are quasi-degenerate, in this work
we take  the normal mass order $m_{\nu 1}<m_{\nu
2}<m_{\nu 3}$ with values:
\footnote{In general,
the impact of RG evolution on neutrino masses and mixing
angles  can be large, however, it is small for a hierarchy of 
neutrinos we chosen~\cite{Casas00}.}
\beq 
m_{\nu 1}&=&0, \ \ \ m_{\nu 2}=\sqrt{\Delta m^2_{sol}},
  \ \ \ m_{\nu 3}=\sqrt{\Delta m^2_{atm}} . 
\eeq 
The mixing angles are fixed to be 
\beq
\theta_{L12}&=&\theta_{sol}=30^0,\ \ \
\theta_{L23}=\theta_{atm}=45^0 . 
\eeq 
Further, we restrict $\theta_{L13}<10^0$. 
Then $(\yndyn)_{IJ}$ in Eq.\ (\ref{yy}) are given by 
\beq 
(\yndyn)_{12}&\simeq& \frac{\sqrt{2}}{4v^2\sin^2\beta}\left(\frac{\sqrt{3}}{2}
m_2^2+\sin 2\theta_{L13} m_3^2\right), \\
(\yndyn)_{13}&\simeq&\frac{\sqrt{2}}{4v^2\sin^2\beta}\left(-\frac{\sqrt{3}}{2}
m_2^2+\sin
2\theta_{L13} m_3^2\right), \\
(\yndyn)_{23}&\simeq&\frac{1}{4v^2\sin^2\beta}\left(2
m_2^2-m_3^2\right). 
\eeq 
The dependence of the parameter $\yndyn$
on CP phase $\delta_L$ is very weak and thus has been neglected.

\begin{figure}[htb]
\epsfig{file=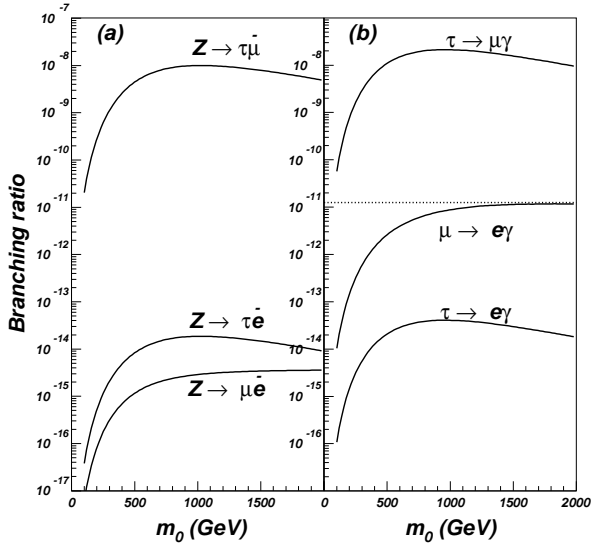,width=8cm}
\caption{Branching ratios of $Z\to \ell_I\bar{\ell}_J$ and
         $\ell_J\to\ell_I\gamma$ versus the common scalar mass $m_0$.
Other parameters are fixed to be  $m_{1/2}=800\ GeV$, $A_0=0$, 
$\tan\beta=10$, $m_2=10$ GeV, $m_3/m_2=30$ and $\theta_{L13}=0$. 
The dash line in (b) is the experimental upper bound on $\mu\to e\gamma$.
\label{Fig:Fig1}}
\end{figure}
{\em (2) Experimental bounds on LFV lepton decays:~}\\
LFV lepton decays have been searched in
several experiments and the current bounds are given by
\cite{Brooks99,Edwards96,Akers95,Abreu97}
\beq
{\cal B}r(\mu\to e\gamma)&<&1.2\times 10^{-11}, \\
{\cal B}r(\tau\rightarrow (e,\ \mu)\gamma)&<&(2.7,\ 1.1)\times 10^{-6},
\label{brL} \\
{\cal B}r(Z\to\tau \bar{\mu})&<&1.2\times 10^{-5}, \\
{\cal B}r(Z\to (\mu,\ \tau)\bar{e})&<&(1.7,\ 9.8)\times 10^{-6}.
\label{brZ}
\eeq

In addition, the explanation of the observed lepton number
asymmetry by seesaw mechanism gives a lower bound for heavy
Majorana neutrinos ${\cal M}_1> 10^{11}\ GeV$ \cite{Endoh02}.
Taking into the constraints mentioned above and fixing the
right-handed neutrino masses as ${\cal M}_1=10^{13}\ GeV$,  ${\cal
M}_2\simeq 10^{15}\ GeV$, we solve the full RG equations listed in
Appendix A numerically based on the work of \cite{Djouadi02},
where the experimental bounds from $b\to s\gamma$ and $g_\mu-2$
have been already taken into account. Although the processes
$Z\to\ell_I\bar{\ell}_J$ are closely correlate to
$\ell_J\to\ell_I\gamma$ and there is a quite stringent bound on
$\mu\to e\gamma$, our numerical results show that there exists  a
scenario   with $m_2\ll m_3$ and a very small $\theta_{L13}$, in
which a large branching ratio for $Z\to\tau\bar{\mu}$ is obtained.

In Fig.~\ref{Fig:Fig1} we show the branching ratios
of $Z\to \ell_I\bar{\ell}_J$ and $\ell_J\to\ell_I\gamma$
versus the common scalar mass $m_0$.
From Fig.~\ref{Fig:Fig1} we have the following observations:
\begin{itemize}
\item[{\rm (1)}]  With fixed $m_{1/2}$ and $\tan\beta$, both
${\cal B}r(Z\to\tau \bar{\mu})$ and ${\cal B}r(\tau \to \mu\gamma)$
reach their maximum values as $m_0\simeq 1000$ GeV, then drop slowly
as  $m_0$ gets larger.
\item[{\rm (2)}] The branching ratio of $Z\to \tau\bar{\mu}$ can be as large as
    $10^{-8}$.
\end{itemize}
    Since $5.5\times 10^9$ $Z$-bosons will be produced
    at the LHC \cite{Martin00} and the possible sensitivity of
    GigaZ to $Z\to \tau\bar{\mu}$ will be up to $10^{-8}$ \cite{AS01},
    the mode $Z\to \tau\bar{\mu}$  will be accessible at both the LHC
    and TESLA GigaZ.
\begin{figure}[htb]
\epsfig{file=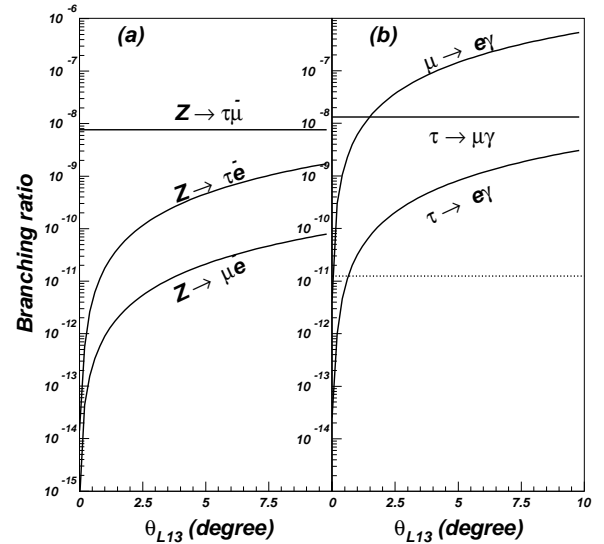,width=8cm}
\caption{ Same as Fig. \ref{Fig:Fig1}, but versus the mixing angle $\theta_{L13}$
          with $m_0=500\ GeV$.}
\label{Fig:Fig2}
\end{figure}
It is noticeable that the branching ratios are sensitive to
the mixing angle $\theta_{L13}$
except for the processes $Z\to\tau\bar{\mu}$ and $\tau\to \mu\gamma$.
As an illustration, we plot the dependence on $\theta_{L13}$
in Fig.~\ref{Fig:Fig2}.
We see that to satisfy the experimental constraint
on $\mu\to e\gamma$, the mixing angle $\theta_{L13}$ must
be quite small. Therefore, a joint measurements for
LFV $Z$-decays and lepton decays will set strong constraints
on the model parameter space.

\section{Conclusions}

We evaluated the lepton flavor violation $Z$ decays in the framework of 
supersymmetric seesaw model at first time.
Although different forms of neutrino couplings may lead to 
different size of LFV $Z$ decays,  we emphasize that it is 
important to study how large the rate for the LFV can be for some 
typical cases and analyze the possibility to observe 
$Z\rightarrow \ell_I\bar{\ell}_J$ in future experiments.
From our calculation results we conclude that,
subject to the constraints from the existing neutrino oscillation
data and the experimental bounds on the decays $\ell_J\to\ell_I\gamma$,
the LFV $Z$-decays $Z\to\ell_I\bar \ell_J$
can still be sizable in supersymmetric seesaw model,
among which the largest-rate channel $Z\to \tau \bar{\mu}$ can occur
with a branching ratio of $10^{-8}$ and thus may be
accessible at the LHC and GigaZ experiment.

\section*{Acknowledgements}

We thank T. Morozumi, J.I. Illnan 
and C. Grosche for very useful discussions and comments.
The work of Z. Xiong was supported  by the Grant-in-Aid for 
JSPS Fellows (No.1400230400).

\appendix
\section{ Renormalization group equations in SUSY seesaw model}

In this appendix we present $additional$ contributions to 
the RG equations of some parameters in supersymmetric seesaw model
due to non-zero neutrino interactions.
The detailed description of these equations
can be found in \cite{Casas01,Hisano02}.
At one-loop level, the RG equations are given as follows.

(1) For Yukawa couplings:
\beq
\frac{d\y_\nu}{dt}&=&\frac{\y_\nu}{16\pi^2}
              \left(T_2-g_1^2-3g_2^2+3\yndyn+\yldyl\right),\\
\frac{d\y_\ell}{dt}&=&\frac{\y_\ell}{16\pi^2}\yndyn,\\
\frac{d \y_U}{dt}&=&\frac{\y_U}{16\pi^2}Tr(\yndyn),
\eeq
where $t=\ln\mu_r$ with $\mu_r$ being the renormalization scale,
and $T_2=Tr(3\yudyu+\yndyn)$. $\y_U$ is the Yukawa coupling
matrix for up-type quarks,  and $g_1$, $g_2$ and $g_3$
are the $U(1)_Y$, $SU(2)$ and $SU(3)$ gauge coupling constants,
respectively.

(2) For soft parameters:
\beq
\frac{d\m_\tl^2}{dt}&=&\frac{1}{16\pi^2}\left[
        \m_\tl^2\yndyn+\yndyn\m_\tl^2+
          2\y_\nu^\dag\m^2_\tn\y_\nu\right.\non
&&\left.+2m_{H_2}^2\y_\nu^\dag\y_\nu
+2\A_\nu^\dag\A_\nu\right],\\
\frac{d\m_\tn^2}{dt}&=&\frac{1}{8\pi^2}\left[
\m^2_\tn\y_\nu\y_\nu^\dag+\y_\nu\y_\nu^\dag\m^2_\tn
+2\y_\nu\m^2_\tl\y_\nu^\dag\right.\non
&&\left.+2m^2_{H_2}\y_\nu\y_\nu^\dag+2\A_\nu\A_\nu^\dag\right],\\
\frac{dm_{H_2}^2}{dt}&=&\frac{1}{8\pi^2}Tr\left[\y_\nu^\dag
                    \left(\m_\tl^2+\m^2_\tn+m_{H_2}^2\right)\y_\nu
\right.\non
&&\left.+\A_\nu^\dag\A_\nu\right],\\
\frac{d\A_\ell}{dt}&=&\frac{1}{16\pi^2}\left(
        2\y_\ell\y_\nu^\dag\A_\nu+\A_\ell\yndyn\right),\\
\frac{d\A_\nu}{dt}&=&\frac{1}{16\pi^2}\left\{\left[T_2-g_1^2-3g_2^2
+4\y_\nu\y_\nu^\dag\right]\A_\nu+\A_\nu\yldyl\right.\non
&&\left.+\left[2Tr(3\y_U^\dag\A_U+\y_\nu^\dag\A_\nu)
+5\A_\nu\y_\nu^\dag\right]\y_\nu\right.\non
&&\left.-2\left(g_1^2M_1+3g_2^2M_2\right)\y_\nu
+2\y_\nu\y_\ell^\dag\A_\ell\right\}.
\eeq

(3) For neutrino masses \cite{Antusch02}:
\beq
\frac{d\M}{dt}&=&\frac{1}{8\pi^2}\left[
\M(\y_\nu\y_\nu^\dag)^T+\y_\nu\y_\nu^\dag\M\right],\\
\frac{d\M_\nu}{dt}&=&\frac{1}{16\pi^2}\left\{
                  \left[2T_2+(\yndyn+\yldyl)^T\right]
                    \M_\nu\right.\non
           &&\left.+\M_\nu\left(\yndyn+\yldyl
                   -2g_1^2-6g_2^2\right)
\right\} .
\label{rge}
\eeq
Note that the above RG equations are valid for the running from $M_P$ to
$M$. Below the scale $M$, the RG equations are the same except
that the couplings of the right-handed neutrinos do not appear.


\end{document}